\documentclass[epj]{svjour}
\usepackage{graphics}
\usepackage{amsmath}
\usepackage{braket}
\usepackage[toc,page]{appendix}
\usepackage{mathtools}
\usepackage{extarrows}
\usepackage{tensor}

\usepackage{ucs}
\usepackage[utf8]{inputenc}
\DeclareMathOperator\arctanh{arctanh}

\newcommand{\be}{\begin{equation}}
\newcommand{\ee}{\end{equation}}
\newcommand{\bea}{\begin{eqnarray}}
\newcommand{\eea}{\end{eqnarray}}

\begin{document}
\title{Spinor wave function of the Universe in non-minimally coupled varying constants cosmologies}

\titlerunning{Spinor wave function of the Universe in the varying constants cosmologies}


\author{Adam Balcerzak\inst{1,2} \and Mateusz Lisaj\inst{3}
% etc
}

\authorrunning{A. Balcerzak, M. Lisaj}
\institute{Institute of Physics, University of Szczecin,
Wielkopolska 15, 70-451 Szczecin,  Poland \and Copernicus Center for Interdisciplinary Studies, Szczepa\'nska 1/5, 31-011 Krak\'ow, Poland \and  
Institute of Mathematics, Physics and Chemistry, Maritime University of Szczecin, Wa{\l }y Chrobrego 1-2, 70-500 Szczecin, Poland}
\date{Received: date / Revised version: date}
\abstract{In this paper, we introduce a non-minimally coupled varying speed of light and varying gravitational constant cosmological toy model. Using the Eisenhart-Duval lifting method, we extend the original minisuperspace of the model and depict the evolution of the system in the presence of the potential term as a geometrical flow associated with the lifted metric. We write the Dirac-Wheeler-DeWitt equation, which solution is a spinor wave function of the Universe. Then we find the solution of the Dirac-Wheeler-DeWitt equation, which describes the emergence of two early universe-antiuniverse pairs that differ with the conserved quantity, which is an analog of the spin.
\PACS{
      {04.50.Kd}{Modified theories of gravity}   \and
      {04.60.-m}{Quantum gravity}
     } % end of PACS codes
} %end of abstract
\maketitle

\section{Introduction}
\label{intro}
In \cite{Kan1,Kan2}, it was shown that the extension of minisuperspace using the Eisenhart-Duval lift \cite{Eisenhart,Duval,Finn} provides a natural framework for introducing a Dirac-type equation. This equation can replace the Wheeler-DeWitt equation, which, as a hyperbolic partial differential equation, does not necessarily provide a positively defined probability density. This issue was addressed by introducing a Dirac-square root of the Wheeler-DeWitt equation  \cite{Mallett,Kim,Death,Yamazaki} which leads to ambiguities related to factor ordering, or by using supersymmetric quantum mechanics \cite{Hojman,moniz1,moniz2,moniz3,moniz4,moniz5}. The crucial concept in the Eisenhart-Duval lift is the notion of an extended minisuperspace, which is obtained by adding an auxiliary dynamical variable to the original set of variables defining the initial minisuperspace associated with the considered dynamical system. This auxiliary variable parametrizes an additional dimension of the extended minisuperspace. In such an extended minisuperspace, the dynamics of the system, which generally is not a geodesic one in the presence of the potential term, can be transformed into a geodesic dynamics associated with the so-called lifted metric that is suitably defined on the extended minisuperspace, since the ``extended’’ equation of motion is expressed by the Laplace-Beltrami operator and thus exhibits covariance. Moreover, in \cite{Kan1,Kan2} it was argued that the requirement of the covariance in the extended minisuperspace resolves the problem of factor ordering and may provide a consistent way to formulate the Dirac square root.

Most attempts to incorporate varying speed of light into cosmological models result in a violation of Lorentz invariance \cite{Albrecht,Barrow1,Magueijo1,Clayton,Drummond,Clayton2}. Such theories also violate general covariance, which requires a preferred reference frame, usually identified as the cosmological frame. Nevertheless, some solutions have been proposed that address the horizon, flatness, and cosmological constant problem \cite{Albrecht,Barrow1}. However, these solutions come at the cost of postulating  ad hoc the time dependence of the speed of light, as the governing terms of the speed of light dynamics are lacking in the action.

The varying speed of light and varying gravitational constant cosmological model considered in this paper is based on the model proposed in \cite{Magueijo1} that includes covariance and the local Lorentz invariance in the case which allows for variation of the speed of light. In such a model, both the speed of light and the gravitational constant are additional dynamic degrees of freedom represented by the two scalar fields with appropriate kinetic terms in the action. It was shown that such model includes a scenario in which the Universe is created in the process which is an analog of quantum scattering on the potential barrier \cite{Balcerzak1} (this is similar to what was found in \cite{Veneziano}), that the third quantization of the model leads to a scenario of the emergence of the multiverse from nothing \cite{Balcerzak2}, and that the created pairs of universes are entangled which is reflected by non-vanishing entanglement entropy \cite{Balcerzak3,Balcerzak4}. Third quantization of slightly different models, in which the fundamental constants undergo cyclic evolution \cite {Marosek}, yields insight into the relationship between quantum entanglement and thermodynamics \cite{Robles_Bal}.

As discussed earlier, the solutions of Wheeler-DeWitt equation which describes the canonically quantized model of varying speed of light and gravitational constant,  include interesting cosmogenesis scenarios that are similar to those found in string cosmological models \cite{Veneziano,Buonanno,Gasperini}. In this paper, we will investigate the potential implications of incorporating spinorial properties of the wave function as a solution to the Dirac-like equation in the quantized varying fundamental constants cosmological model, as such an approach may lead to some novel features of quantum cosmogenesis.

Our paper is organized as follows. In Sec. \ref{sec:1} we describe the non-minimally coupled varying speed of light $c$ and varying gravitational constant $G$ cosmological model and provide the description of its high-curvature limit. In Sec. \ref{sec:2} we follow the Eisenhart-Duval lifting scheme and formulate the dynamics of our model on the extended minisuperspasce. Then, utilizing the geometric nature of the dynamics of our "lifted" system, we formulate the Dirac-Wheeler-DeWitt equation.  In Sec. \ref{sec:3} we obtain the solution to the Dirac-Wheeler-DeWitt equation and establish its connection to the scenario of cosmogenesis. In Sec. \ref{sec:conc} we give our conclusions.

\section{The non-minimally coupled varying $c$ and $G$ cosmological model}
\label{sec:1}
\begin{figure}
\begin{center}
\resizebox{0.4\textwidth}{!}{\includegraphics{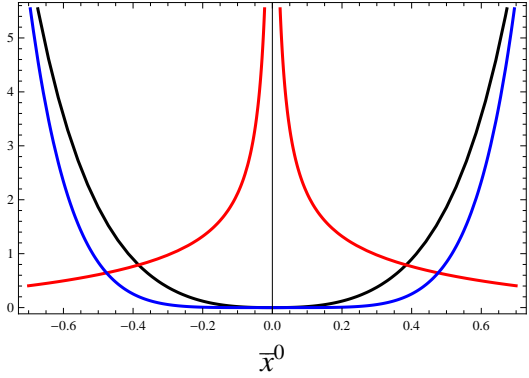}}
\caption{\label{acg} The evolution of the scale factor $a$ (in black), the speed of light $c$ (in red), and the gravitational constant $G$ (in blue) over time is presented for $\bar{x}^0 < 0$ and $\bar{x}^0 > 0$, corresponding to the phase before and after the curvature singularity, respectively.}
\end{center}
\end{figure}

Our model, which proposes variations in the speed of light $c$ and the gravitational constant $G$, is primarily represented by the theory of non-minimally coupled bi-scalar gravity. In this theory, the values of both fundamental constants are connected to the values of the two scalar degrees of freedom \cite{Balcerzak1}. The model under consideration is based on the covariant and locally Lorentz-invariant theory of varying speed of light, as described in \cite{Magueijo1}. It is defined by the following action:
\begin{equation}
\label{action}
S=\int \sqrt{-g}  \left(\frac{e^{\phi}}{e^{\psi}}\right) \left[R+\Lambda + \omega (\partial_\mu \phi \partial^\mu \phi + \partial_\mu \psi \partial^\mu \psi)\right] d^4x,
\end{equation}
where $\phi$ and $\psi$ represent non-minimally coupled scalar fields, $R$ denotes the Ricci scalar, $\Lambda$ acts as the cosmological constant, and $\omega$ is a parameter of the model. The action specified in equation (\ref{action}) was derived from the original Einstein-Hilbert action by substituting the functions of the scalar fields $\phi$ and $\psi$ for the speed of light $c$ and the gravitational constant $G$. The exact connection between the scalar fields $\phi$ and $\psi$ and the constants $c$ and $G$ is given by the following formulas:
\bea
c^3&=&e^{\phi}, \\
G&=&e^\psi.
\eea
The application of the following fields transformations:
\begin{eqnarray}
\label{fred}
\phi &=& \frac{\beta}{\sqrt{2\omega}}+\frac{1}{2} \ln \delta, \\
\psi &=& \frac{\beta}{\sqrt{2\omega}}-\frac{1}{2} \ln \delta,
\end{eqnarray}
causes the original action (\ref{action}) to take the form of the Brans-Dicke action given by the following expresion:
\bea
\label{actionBD} \nonumber
S=\int \sqrt{-g}\left[ \delta (R+\Lambda) +\frac{\omega}{2}\frac{\partial_\mu \delta \partial^\mu \delta}{\delta} + \delta \partial_\mu \beta \partial^\mu \beta\right]d^4x.\\
\eea
The dependence of the speed of light on the space and time variables causes a violation of the general covariance in our model, which is a common characteristic of most theories that propose a variable speed of light. To address this, a specific frame of reference, known as the ''light frame'', must be selected. This frame serves as the preferred reference frame for the formulation of the model. Based on the proposal, outlined in \cite{Albrecht,Barrow1}, we identify the light frame for our model with the cosmological frame defined by the FLRW metric:
\be
\label{FLRW}
ds^2=-N^2(dx^0)^2+ a^2(dr^2 + r^2 d\Omega^2),
\ee
where $N$ is the lapse function and $a$ is the scale factor, both dependent on the coordinate $x^0$. The insertion of the metric from equation (\ref{FLRW})  into equation (\ref{actionBD}) gives the following form of the action of our model in the cosmological frame:
\begin{eqnarray}
\label{action_sym} \nonumber
S &=& \frac{3 V_0}{8 \pi} \int dx^0 \left(-\frac{a^2}{N} a' \delta' - \frac{\delta}{N} a a'^2   + \Lambda \delta a^3 N  \right. \\
&-& \left.\frac{\omega}{2} \frac{a^3}{N} \frac{\delta'^2}{\delta}-\frac{a^3}{N}\delta \beta'^2 \right),
\end{eqnarray}
where $()'\equiv \frac{\partial}{\partial x^0}$. The solution to the model described by action (\ref{action_sym}) expressed in the gauge defined by $N = a^3 \delta$ is \cite{Balcerzak1}:
\bea
\label{rozwio1}
a&=& \frac{1}{D^2 {(e^{ F x^0})}^2 \sinh ^ M |\sqrt{(A^2-9)\Lambda }x^0| },\\
\label{rozwio2}
\delta &=& \frac{D^6 {(e^{ F x^0})}^6}{\sinh ^ W |\sqrt{(A^2-9)\Lambda }x^0|},
\eea
where   $A=\frac{1}{\sqrt{1-2\omega}}$, $M=\frac{3-A^2}{9-A^2}$, $W=\frac{2A^2}{9-A^2}$ while $D$ and $F$ are the integration constants. The relation between $x^0$ and the rescaled cosmic time $\bar{x}^0$ is given by \cite{Balcerzak1}:
\begin{equation}
\label{conect}
\begin{split}
x^0 &= \frac{2}{\sqrt{(A^2-9)\Lambda}}  \arctanh \left(e^{\sqrt{(A^2-9)\Lambda}\bar{x}^0}\right)\,,
\hspace{0.2cm}
\text{for $\bar{x}^0<0$}\,,
\\
x^0 &= \frac{2}{\sqrt{(A^2-9)\Lambda}}  \arctanh \left(e^{- \sqrt{(A^2-9)\Lambda}\bar{x}^0}\right)\,,
\hspace{0.2cm}
\text{for $\bar{x}^0>0$}\,,
\end{split}
\end{equation}
where as in \cite{Balcerzak1} we restrict the considered class of models to the cases with $A^2>9$. The pre-big-bang contraction occurring for $\bar{x}^0<0, $ followed by the post-big-bang expansion taking place for $\bar{x}^0>0$ are both described by the solution given by (\ref{rozwio1}) and (\ref{rozwio2}), along with (\ref{conect}). The two phases are separated by the curvature singularity at $\bar{x}^0=0$. The evolution of the fundamental constants $c$ and $G$ is also conveyed in formulas (\ref{rozwio1}) and (\ref{rozwio2}). According to these formulas, the gravitational constant $G$ tends to zero while the speed of light $c$ becomes infinite as the universe approaches the curvature singularity located at $\bar{x}^0=0$ (see Fig. (\ref{acg})).

By utilizing the field transformations given by:
\bea
\nonumber
X = \ln(a \sqrt{\delta}),&& \hspace {0.3cm} Y = \frac{1}{2A} \ln \delta,\\ \nonumber
\eta=r(AY-3X), \hspace {0.1cm} x_1&=&r(3Y-AX), \hspace {0.1cm} x_2=2\sqrt{\tilde{V}_0}\beta,\\
\eea
where $\tilde{V}_0 = \frac{3 V_0}{8\pi}$ and $r=2\sqrt{\frac{\tilde{V}_0}{A^2-9}}$ the action (\ref{action_sym}) can be transformed to the following form:
\be
\label{action_simple}
S= \int dx^0 \left[\frac{1}{4}(\eta'^2-x_1'^2-x_2'^2)+\bar{\Lambda}e^{-2\frac{\eta}{r}}\right],
\ee
where $\bar{\Lambda} = \tilde{V}_0\Lambda$. The Hamiltonian associated with the action (\ref{action_simple}) reads:
\be
\label{ham}
H=\pi_\eta^2 -\pi_{x_1}^2- \pi_{x_2}^2-\bar{\Lambda}e^{-2\frac{\eta}{r}},
\ee
where $\pi_\eta=\frac{\eta'}{2}$, $\pi_{x_1}=-\frac{x_1'}{2}$ and $\pi_{x_2}=-\frac{x_2'}{2}$ are the conjugated momenta. From the structure of the Hamiltonian (\ref{ham}), it can be inferred that both $\pi_{x_1}$ and $\pi_{x_2}$ remain conserved throughout the evolution. Consequently, the classical evolution can be seen as a particle scattering off the exponential potential barrier. The solutions to the Hamilton's equations associated with the Hamiltonian (\ref{ham}) are as follows:
\bea
\label{ham_sol1}
\eta&=&\ln \sinh|\sqrt{(A^2-9)\Lambda }x^0|, \\
\label{ham_sol2}
x_1&=& -2 \pi_{x_1} x^0 + E, \\
\label{ham_sol3}
x_2&=&-2 \pi_{x_2} x^0 + P,
\eea
where $E$ and $P$ are integration constants. Upon inspection of the solution (\ref{ham_sol1}), it is evident that $\eta$ can define two distinct regimes. The first one is the high-curvature regime, which is characterized by the vanishing scale factor $a\rightarrow 0$ and corresponds to $\eta\rightarrow \infty$. The second one is the low-curvature regime, which is defined by higher values of the scale factor $a$ and occurs for $\eta\rightarrow -\infty$. Furthermore, it can be verified that the high-curvature regime, specifically when $\eta\rightarrow \infty$, is characterized by the following asymptotic values of momentum $\pi_\eta$: 
\[ \pi_\eta = \left\{
  \begin{array}{l l}
    \sqrt{\bar{\Lambda}} & \quad \text{(collapsing pre-big-bang solution)}\\
    -\sqrt{\bar{\Lambda}} & \quad \text{(expanding post-big-bang solution).}
  \end{array} \right.\]
 As for the low-curvature regime (as $\eta\rightarrow -\infty$), we can observe that:
    \[ \pi_\eta = \left\{
  \begin{array}{l l}
    \sqrt{\bar{\Lambda}} e^{-\frac{\eta}{r}} & \quad \text{(collapsing pre-big-bang solution)}\\
    -\sqrt{\bar{\Lambda}} e^{-\frac{\eta}{r}} & \quad \text{(expanding post-big-bang solution).}
  \end{array} \right.\]
  
 To derive the Wheeler-DeWitt equation that characterizes the quantum mechanical aspect of the model, we implement the Jordan quantization rules which involve replacing the canonical momenta with the corresponding operators according to the following scheme: $\pi_\eta\rightarrow \hat{\pi}_\eta=-i \frac{\partial}{\partial \eta}$, $\pi_{x_1}\rightarrow \hat{\pi}_{x_1}=-i \frac{\partial}{\partial x_1}$ and $\pi_{x_2}\rightarrow \hat{\pi}_{x_2}= -i \frac{\partial}{\partial x_2}$. The resulting Wheeler-DeWitt equation can be expressed as:
 \begin{equation}
\label{KG}
\ddot{\Phi} - \Delta \Phi + m_{eff}^2(\eta) \Phi=0,
\end{equation}
where $\dot{( )}\equiv\frac{\partial}{\partial \eta}$, $\Delta = \frac{\partial^2}{\partial x_1^2}+\frac{\partial^2}{\partial x_2^2}$ and $m_{eff}^2(\eta)= \bar{\Lambda} e^{-\frac{2}{r}\eta}$.

\section{Quantization of the non-minimally coupled varying $c$ and $G$ cosmological model. The Dirac-Wheeler-DeWitt equation}
\label{sec:2}

We will be following the Eisenhart-Duval lifting scheme \cite{Eisenhart,Duval} adjusted to the cosmological setup based on the scalar-tensor gravity \cite{Kan1,Kan2,Finn}. In this method the minisuperspace associated with considered model is extended by introducing additional dynamical variable which makes possible to describe the evolution of the system in purely geometrical terms even in the presence of the potential. In other words the nongeometrical evolution can be depicted as a geodesic flow in the extended minisuperspace, that respects the lifted metric (the metric on the extended minisuperspace) since the Hamiltonian of the lifted system is represented by the Laplacian. The starting point is the Lagrangian defined by the action (\ref{action_simple}) which reads as: 
\be
\label{lag}
L=\frac{1}{4}(\eta'^2-x_1'^2-x_2'^2)+\bar{\Lambda}e^{-2\frac{\eta}{r}}.
\ee
The lifted Lagrangian associated with (\ref{lag}) is the following \cite{Kan1}:
\be
\label{lag_ext}
L_{ext}= \frac{1}{2}\left( \frac{\eta'^2}{2} - \frac{x_1'^2}{2} - \frac{x_2'^2}{2} - \frac{x_3'^2}{2 \bar{\Lambda}e^{-2\frac{\eta}{r}}}\right),
\ee
where $x_3$ parametrizes an additional degree of freedom of the extended minisuperpace. The metric of the extended minisuperspace (the lifted metric) is then given by \cite{Kan1}:
\be
\label{metric_ext}
\tilde{G}_{\alpha\beta}=diag\left(\frac{1}{2},-\frac{1}{2},-\frac{1}{2}, \frac{-1}{2 \bar{\Lambda}e^{-2\frac{\eta}{r}}}\right).
\ee
Since, the Hamiltonian constraint
\be
\label{ham_const}
\frac{1}{2} \tilde{G}^{\alpha\beta} \tilde{P}_\alpha \tilde{P}_\beta=0,
\ee
where $\tilde{P}_\alpha=\tilde{G}_{\alpha\beta} x'^\beta$, is conformally invariant, namely, it is essentially the same as $\tilde{G}^{\alpha\beta} \rightarrow \Omega^2 \tilde{G}^{\alpha\beta}$, we will be using the following conformally equivalent extended minisuperspace metric
\be
\label{metric_ext_conf}
G_{\alpha\beta}=\Omega^2 \tilde{G}_{\alpha\beta},
\ee
with $\Omega^2=\left[{2 \bar{\Lambda}e^{-2\frac{\eta}{r}}}\right]^\frac{1}{n-2},$ where $n$ stands for the dimension of the extended minisuperspace. Since in our case $n=4$ the conformal factor is given by:
\be
\label{conf_fact}
\Omega^2=\sqrt{2 \bar{\Lambda}}e^{-\frac{\eta}{r}}.
\ee
Thus, the conformally equivalent extended minisuperspace metric is the following:
\be
\label{the_metric}
G_{\alpha\beta}=diag\left(\sqrt{\frac{\bar{\Lambda}}{2}}e^{-\frac{\eta}{r}},-\sqrt{\frac{\bar{\Lambda}}{2}}e^{-\frac{\eta}{r}},-\sqrt{\frac{\bar{\Lambda}}{2}}e^{-\frac{\eta}{r}},\frac{-e^{\frac{\eta}{r}}}{\sqrt{2\bar{\Lambda}}}\right).
\ee
The covariant Wheeler-DeWitt equation for the extended case is given by:
\be
\label{WDW_cov}
\frac{1}{\sqrt{-G}}\partial_\alpha (\sqrt{-G} G^{\alpha\beta} \partial_\beta \Phi)=0,
\ee
where $G$ is the determinant of $G_{\alpha\beta}$. Substitution of (\ref{the_metric}) into (\ref{WDW_cov}) leads to the following Wheeler-DeWitt equation for the extended minisuperspace:
\be
\label{WDW_ext}
\partial_\eta^2 \Phi - \partial_{x_1}^2 \Phi - \partial_{x_2}^2 \Phi - \bar{\Lambda} e^{-2\frac{\eta}{r}} \partial_{x_3}^2 \Phi=0.
 \ee
If we impose the condition 
\be
\label{reduce}
\partial_{x_3}^2 \Phi=-\Phi,
\ee
the Wheeler-DeWitt equation for the extended minisuperspace (\ref{WDW_ext}) reduces to the initial one (\ref{KG}).

As in the case of Wheeler-DeWitt equation we will use the covariance in the extended minisuperspace as guiding principle which leads to the proper form of the Dirac-Wheeler-DeWitt equation \cite{Kan1,Kan2}. Since the Dirac equation without the mass term has the conformal covariance \cite{Hijazi} we will again use (\ref{the_metric}) as the lifted metric. Thus, the covariant Dirac-Wheeler-DeWitt equation is:
\be
\label{dirac} 
\hat{\gamma}^\alpha D_\alpha \Psi \equiv \gamma^A \tensor{e}{_A ^\alpha} D_\alpha \Psi=0,
\ee
 where $\gamma^A$ are Dirac matrices:
\begin{eqnarray}
\gamma^{0} =
\begin{pmatrix}
I & 0  \\
0 & -I \\
\end{pmatrix}, \hspace{0.5 cm}
\gamma^{k} =
\begin{pmatrix}
0 & \sigma_k \\
-\sigma_k & 0 \\
\end{pmatrix},
\end{eqnarray}
with $k=1,2,3$, $\sigma_k$ being Pauli matrices and $I$ being an identity matrix. The coefficients $\tensor{e}{_A ^\alpha}$ in (\ref{dirac}) are four vector fields defined by the expression $\eta_{AB}=\tensor{e}{_A ^\alpha} \tensor{e}{_B ^\beta}  G_{\alpha\beta}$ and $D_\alpha$ is a covariant derivative given by:
\be
\label{cov_der}
D_\alpha=\partial_\alpha + \Gamma_\alpha,
\ee
where 
\be
\label{big_gamma}
\Gamma_\alpha=\frac{1}{2} \omega_{AB_\alpha} \Sigma^{AB}
\ee
with the spin connection $\omega_{AB_\alpha}$  defined as:
\be
\label{omega}
\omega_{AB_\alpha} = G_{\nu\mu} \tensor{e}{_A ^\mu} \nabla_\alpha \tensor{e}{_B ^\nu} 
\ee
and
\be
\label{big_sigma}
\Sigma^{AB}=\frac{1}{4}\left[\gamma^A,\gamma^B \right].
\ee
The explicit form of the vector field $\tensor{e}{_A ^\alpha} $ is given by:
\be
\label{vector_fields}
\tensor{e}{_A ^\alpha} =diag \left(\left(\frac{\bar\Lambda}{2} \right)^{-\frac{1}{4}}e^{\frac{\eta}{2r}}\left(1,1,1\right), \left(2 \bar\Lambda\right)^{\frac{1}{4}}e^{-\frac{\eta}{2r}}\right)
\ee
and the non-vanishing elements of the spin connection $\omega_{AB_\alpha}$ are:
\begin{eqnarray}
\label{spin_conn}
\omega_{10_1}&=&\frac{1}{2r}, \hspace{0.5cm} \omega_{01_1}=-\frac{1}{2r}, \nonumber \\
\omega_{20_2}&=&\frac{1}{2r}, \hspace{0.5cm} \omega_{02_2}=-\frac{1}{2r},  \\
\omega_{30_3}&=&\frac{-e^{\frac{\eta}{r}}}{2r\sqrt{\bar\Lambda}}, \hspace{0.5cm} \omega_{03_3}=\frac{e^{\frac{\eta}{r}}}{2r\sqrt{\bar\Lambda}}. \nonumber
\end{eqnarray}
The explicit form of the Dirac-Wheeler-DeWitt equation in the extended minisuperspace is then given by:
\bea
\label{dirac_expl}
\nonumber
\left[\gamma^0 \left(\frac{\partial}{\partial \eta} -\frac{1}{4r}\right) +\gamma^1 \frac{\partial}{\partial x_1}  + \gamma^2 \frac{\partial}{\partial x_2}  + \sqrt{\bar\Lambda} e^{-\frac{\eta}{r}}\gamma^3 \frac{\partial}{\partial x_3} \right] \Psi =0.\\
\eea
The reduction of the extended Dirac-Wheeler-DeWitt equation (\ref{dirac_expl}) to the initial minisuperspace can be obtained by the application of the assumption which reads:
\be
\label{reduction}
\frac{1}{i} \frac{\partial}{\partial x_3} \Psi=\Psi.
\ee

\section{Spinor wave function of the Universe in the non-minimally coupled varying $c$ and $G$ model}
\label{sec:3}

The solution of the Dirac-Wheeler-DeWitt equation (\ref{dirac_expl}) reduced with condition (\ref{reduction}) to the initial minisuperspace   is:
\be
\label{spinor}
\Psi = \frac{1}{\sqrt{2}}
\begin{pmatrix}
\phi_1+\varphi_1 \\
\phi_2+\varphi_2 \\
\phi_1-\varphi_1 \\
\phi_2-\varphi_2\\
\end{pmatrix}, 
\ee
with
\begin{flalign}
\nonumber
\phi_1 =&e^{i \vec{k} \cdot \vec{x}} e^{-i \alpha} \alpha^{i k r -\frac{1}{4}}\left(C_1 U(ikr+1, 2ikr+1, 2 i \alpha) \right. &\\
&\left.  +C_2 L^{2 i k r}_{-ikr-1}(2i\alpha)\right), &\\
\nonumber
\phi_2 =&e^{i \vec{k} \cdot \vec{x}} e^{-i \alpha} \alpha^{i k r -\frac{1}{4}}\left(C_3 U(ikr, 2ikr+1, 2 i \alpha) \right. &\\
&\left. +C_4 L^{2 i k r}_{-ikr}(2i\alpha)\right), &\\
\nonumber
\varphi_1 =&e^{i \vec{k} \cdot \vec{x}} e^{-i \alpha} \alpha^{i k r -\frac{1}{4}}\left(C_5 U(ikr, 2ikr+1, 2 i \alpha) \right. &\\
&\left. +C_6 L^{2 i k r}_{-ikr}(2i\alpha)\right), &\\
\nonumber
\varphi_2=&e^{i \vec{k} \cdot \vec{x}} e^{-i \alpha} \alpha^{i k r -\frac{1}{4}}\left(C_7 U(ikr+1, 2ikr+1, 2 i \alpha) \right. &\\
&\left. + C_8 L^{2 i k r}_{-ikr-1}(2i\alpha)\right),
\end{flalign}
where $\vec{k}=(k_1,k_2)$, $\vec{x}=(x_1,x_2)$, $k=\sqrt{k_1^2+k_2^2}$, $U$ and $L$ are the confluent hypergeometric function of the second kind and the associated Laguerre polynomial, respectively, $\alpha=r\sqrt{\bar\Lambda}e^{-\frac{\eta}{r}}$ and $\{C_1, C_2, C_3, C_4, C_5, C_6, C_7, C_8\}$ denotes a set of integration constants.

We introduce a new representation in which the solution (\ref{spinor}) takes simpler form. The unitary matrix that defines such a representation is:
\be
\label{fw}
F= \frac{1}{\sqrt{2}}
\begin{pmatrix}
I & I  \\
I & -I \\
\end{pmatrix},
\ee
and the transformed wave function (\ref{spinor}) is given by:
\be
\label{spinor_fw}
\Psi_F= F \Psi=
\begin{pmatrix}
\phi_1 \\
\phi_2\\
\varphi_1\\
\varphi_2\\
\end{pmatrix}.
\ee
We notice that the unitary transformation given by (\ref{fw}) leaves the four momentum operator unchanged. It also preserves the form of the $x_3$-axis projection spin operator $\Sigma_{x_3}$ given by:
\be
\label{spin_fw}
\Sigma_{x_3}= \frac{1}{2}
\begin{pmatrix}
\sigma_3& 0  \\
0& \sigma_3\\
\end{pmatrix}.
\ee
Let us define now the four spinor wave functions each separately solving the Dirac-Wheeler-DeWitt equation (\ref{dirac_expl}):
\be
\label{pu}
\Psi_{+,+\frac{1}{2}}=A e^{i \vec{k} \cdot \vec{x}} e^{-i \alpha} \alpha^{i k r -\frac{1}{4}}\begin{pmatrix}
L^{2 i k r}_{-ikr-1}(2i\alpha) \\
0\\
U(ikr, 2ikr+1, 2 i \alpha) \\
0\\
\end{pmatrix}, 
\ee
\be
\label{pd}
\Psi_{+,-\frac{1}{2}}=B e^{i \vec{k} \cdot \vec{x}} e^{-i \alpha} \alpha^{i k r -\frac{1}{4}}\begin{pmatrix}
0 \\
U(ikr, 2ikr+1, 2 i \alpha)\\
 0\\
L^{2 i k r}_{-ikr-1}(2i\alpha)\\
\end{pmatrix},
\ee
\be
\label{nu}
\Psi_{-,+\frac{1}{2}}=C e^{i \vec{k} \cdot \vec{x}} e^{-i \alpha} \alpha^{i k r -\frac{1}{4}}\begin{pmatrix}
U(ikr+1, 2ikr+1, 2 i \alpha) \\
0\\
L^{2 i k r}_{-ikr}(2i\alpha)\\
0\\
\end{pmatrix},
\ee
\be
\label{nd}
\Psi_{-,-\frac{1}{2}}=D e^{i \vec{k} \cdot \vec{x}} e^{-i \alpha} \alpha^{i k r -\frac{1}{4}}\begin{pmatrix}
0 \\
L^{2 i k r}_{-ikr}(2i\alpha)\\
 0\\
U(ikr+1, 2ikr+1, 2 i \alpha)\\
\end{pmatrix},
\ee
where $A$, $B$, $C$ and $D$ are some constants.
The spinor wave functions (\ref{pu}), (\ref{nu}) in the low-curvature regime for $\eta\rightarrow-\infty$ ($\alpha\rightarrow\infty$) represent the positive and the negative frequency modes, respectively, with $1/2$ spin along $x_3$ axis, while (\ref{pd}), (\ref{nd}) represent the positive and the negative frequency modes, respectively, with $-1/2$ spin along $x_3$ axis since in this regime:
\be
\label{pos_freq}
i \frac{\partial}{\partial \eta} \Psi_{+,\pm\frac{1}{2}}=\pi_{\eta(-\infty)}\Psi_{+,\pm\frac{1}{2}}
\ee
and
\be
\label{neg_freq}
i \frac{\partial}{\partial \eta} \Psi_{-,\pm\frac{1}{2}}=-\pi_{\eta(-\infty)}\Psi_{-,\pm\frac{1}{2}},
\ee
where $\pi_{\eta(-\infty)}= \sqrt{\bar{\Lambda}} e^{-\frac{\eta}{r}}$.

The general solution of (\ref{dirac_expl}) that holds for any value of the curvature which reads as:
\bea
\label{spinor_general}
\nonumber 
\Psi_{G}&=& \Psi_{+,+\frac{1}{2}} + \Psi_{+,-\frac{1}{2}}+\Psi_{-,+\frac{1}{2}}+\Psi_{-,-\frac{1}{2}} = \\ 
\nonumber
&=&e^{i \vec{k} \cdot \vec{x}} e^{-i \alpha} \alpha^{i k r -\frac{1}{4}}\times \\ 
 \nonumber
&\times&\begin{pmatrix}
A ~ L^{2 i k r}_{-ikr-1}(2i\alpha) + C~ U(ikr+1, 2ikr+1, 2 i \alpha) \\
B ~U(ikr, 2ikr+1, 2 i \alpha)+ D ~L^{2 i k r}_{-ikr}(2i\alpha)\\
A~U(ikr, 2ikr+1, 2 i \alpha)+C~L^{2 i k r}_{-ikr}(2i\alpha) \\
B~L^{2 i k r}_{-ikr-1}(2i\alpha)+D~ U(ikr+1, 2ikr+1, 2 i \alpha)\\
\end{pmatrix}, \\
\eea
represents purely quantum state in the high-curvature limit for $\eta\rightarrow\infty$ since in such limit (\ref{spinor_general}) it is not peaked over any classical trajectory. This can be seen by inspecting (\ref{spinor_general}) which in the limit $\eta\rightarrow\infty$  has the following form:
\bea
\label{spinor_hc}
&&\lim_{\eta \to \infty}\Psi_{G}= e^{i \vec{k} \cdot \vec{x}} e^{-i \alpha} \alpha^{-\frac{1}{4}} \times \nonumber \\ 
&\times&
\begin{pmatrix}
A ~d~ \alpha^{ i k r} + C(a\alpha^{-ikr}+b\alpha^{ikr}) \\
B ~ikr~ (a\alpha^{-ikr}-b\alpha^{ikr}) -D~d~  \alpha^{ikr}\\
A ~ikr~ (a\alpha^{-ikr}-b\alpha^{ikr}) -C~d~  \alpha^{ikr} \\
B ~d~ \alpha^{ i k r} + D(a\alpha^{-ikr}+b\alpha^{ikr})\\
\end{pmatrix},
\eea
where
\bea
a&=&\frac{\Gamma(2ikr)}{ikr \Gamma(ikr)}(2i)^{-2ikr}, \\
b&=&\frac{\Gamma(2ikr)}{-ikr \Gamma(-ikr)}
\eea
and
\bea
d&=&\frac{\Gamma(ikr)}{\Gamma(-ikr)\Gamma(2ikr+1)}.
\eea
The formula (\ref{spinor_general}) shows that in the low-curvature limit for $\eta\to-\infty$ from the pure quantum regime there emerge classical trajectories that represent the expanding post-big-bang universe-antiuniverse pair with the spin projection $\frac{1}{2}$ ($\Psi_{-,+\frac{1}{2}}$ for the universe and $\Psi_{+,+\frac{1}{2}}$ for the antiuniverse) and the expanding post-big-bang universe-antiuniverse pair with the spin projection $-\frac{1}{2}$  ($\Psi_{-,-\frac{1}{2}}$ for the universe and $\Psi_{+,-\frac{1}{2}}$ for the antiuniverse). The scenario discussed here differs from the scenario of cosmogenesis, which involves creation of the Universe via quantum scattering on the exponential potential barrier in the minisuperspace, that appears in the non-minimally coupled varying constants cosmological model, where the ordinary Wheeler-DeWitt equation provides a solution representing the quantum state of the Universe \cite{Balcerzak1}.

\section{Conclusions}
\label{sec:conc}

We have shown that the Eisenhart-Duval lift which is based on the extension of the initial minisuperspace of the model by adding an auxiliary dimension may efficiently be applied in the context of non-minimally coupled varying speed of light and varying gravitational constant model which results in transforming the evolution of the system in the presence of time-dependent mass term into a geodesic evolution compatible with the lifted metric. It should be stressed that the covariance in the extended minisuperspace constitutes a guiding principle that leads to the proper equation of motion \cite{Kan1,Kan2}. In particular we have used this principle in order to construct in a consistent way the Dirac-Wheeler-DeWitt equation which incorporates the spinorial characteristics into the wave function of the Universe. This new quality seems to be particularly interesting since it may substantially influence the distribution when considering an ensemble \cite{Kan2}. We have also shown that there exist the solutions to the Dirac-Wheeler-DeWitt equation that asymptotically, in the low-curvature regime, can be interpreted as the positive and negative frequency modes with the conserved quantity, which is an analog of the spin projection. It should be stressed, however, that asymptotically these solutions exhibit decoupling of the spin analog from the Universe trajectory in the extended minisuperspace which is expected since the analogous situation occurs in case of Dirac equation description of the free electron. 

The considered model is defined by the action which is similar to the low-energy effective action of the string cosmological models  \cite{Veneziano,Buonanno,Gasperini}, therefore, besides the ordinary post-big-bang expanding phase of the evolution it also provides the pre-big-bang cosmological scenario. We have found that the spinor wave function of the Universe which is a solution to the Dirac-Wheeler-DeWitt equation in the high-curvature (near big-bang) regime describes the highly unclassical behavior since it is not peaked over any classical trajectory. On the other hand, in the low-curvature regime (far from the big-bang) the wave function is peaked over classical paths which can be interpreted as a transition from pure quantum behavior to the quasi-classical one and emergence of the two post-big-bang expanding universe-antiuniverse pairs with the opposite spin orientations.

Within the concept of the multiverse, there is a growing approach that regards the minisuperspace as the fundamental arena for physical phenomena to take place \cite{Veneziano,Balcerzak4,Buonanno,Gasperini,Bertolami,Robles_ent1,Robles_ent2,Serrano,Robles1,Kraemer,Barroso3,Barroso4}. Particular models include different types of interactions between the wave functions that represent individual universes \cite{Balcerzak4,Bertolami,Serrano,Robles1,Kraemer,Barroso3,Barroso4}.  This may suggest that a comprehensive inclusion of spin in quantum cosmology requires an application of third quantization procedure \cite{Kan2}, in which the third-quantized action is invariant with respect to some kind of local symmetry.

\appendix

\renewcommand{\theequation}{A.\arabic{equation}}
\renewcommand{\thefigure}{A\arabic{figure}}

\setcounter{equation}{0}
\setcounter{figure}{0}

\begin{appendices}

\section{The Eisenhart-Duval lift, the extended Wheeler-DeWitt  and the Dirac-like equations for a cosmological system}
\label{app:A}
\subsection{The essentials of the Eisenhart-Duval lift}
The Eisenhart-Duval lift \cite{Eisenhart,Duval,Finn} is a formalism that can be used to represent physical systems subjected to conservative forces as an equivalent free system moving on a higher-dimensional curved manifold. Let us start with the Lagrangian for a system of $n$ degrees of freedom (fields):
\begin{equation}\label{Lforce}
\mathcal{L} =\frac{1}{2} k_{ij}(\varphi^1,...,\varphi^n)\;\dot{\varphi}^i\dot{\varphi}^j\: -\: V(\varphi^1,...,\varphi^n)\;,
\end{equation}
where $\varphi^i(t)$ represent a homogeneous scalar field, $ k_{ij}$ is the metric tensor in the configuration space, $ V(\varphi^1,...,\varphi^n)$  represents the potential while the overdot denotes differentiation with respect to $t$. Variation of (\ref{Lforce}) with respect to  $\varphi^i$ gives the following equations of motion:
\begin{equation}\label{EoMforce}
\ddot{\varphi}^i\: +\: \Gamma^i_{jk}\,\dot{\varphi}^j\dot{\varphi}^k\ =\ -\, k^{ij}V_{,j}\;,
\end{equation}
where the symbol $,i$ denotes differentiation with respect to $\varphi^i$, $k^{ij}$ satisfies $k^{il}k_{lj}=\delta^i_j$ and 
\begin{equation}
\Gamma^i_{jk}\ =\frac{1}{2} k^{il}\,\Big(k_{jl,k}+k_{kl,j}-k_{jk,l}\Big).
\end{equation}
We can see that (\ref{EoMforce}) formally represents a geodesic equation of a particle that is subjected to a force.  By utilizing the Eisenhart-Duval lift formalism, we can expand upon this geometric interpretation by generating a manifold of higher dimension for the field space. This manifold will be characterized by trajectories that follow geodesics as described by equation (\ref{EoMforce}) but without any external force. For this purpose we introduce an auxiliary field  $\chi$ and the so called ``lifted'' Lagrangian which reads: 
\begin{equation}\label{Llifted}
{\cal\tilde{ L}}=\frac{1}{2} k_{ij}(\varphi^1,...,\varphi^n)\;\dot{\varphi}^i\dot{\varphi}^j\: +\:\frac{1}{2} \frac{M^4}{V(\varphi^1,...,\varphi^n)}\,\dot{\chi}^2\;,
\end{equation}
where $M$ denotes an arbitrary constant. Such a ``lifted'' Lagrangian can be written as follows:
\begin{equation}\label{LG}
{\cal\tilde{ L}}=\frac{1}{2} G_{AB}\dot{\phi}^A\dot{\phi}^B,
\end{equation}
where $\phi^A=\{\varphi^i,\chi\}$ and 
\begin{equation}
G_{AB}\ =\ 
\begin{pmatrix}  
k_{ij}	&	0\\
0		&	\dfrac{M^4}{V}
\end{pmatrix}.
\end{equation}
We notice that the Lagrangian (\ref{LG}) is identical to the Lagrangian of a free particle moving in a curved spacetime equipped with metric $G_{AB}$ which means that the trajectories of the system are identical to the geodesics of the extended manifold. It can be shown that the geodesics of the extended manifold correspond to the trajectories of the original system described by equations (\ref{EoMforce}). A detailed proof of this correspondence can be found in \cite{Finn}.

\subsection{The extended Wheeler-DeWitt equation for a homogeneous and isotropic Universe}
The Eisenhart-Duval method can be extended to the quantum level described by the Wheeler-DeWitt equation, at least for a homogeneous and isotropic model of the Universe with a single scalar field defined by the following action \cite{Kan1}:
\begin{equation}
S=\int d^4x\sqrt{-g}\left[\frac{1}{2\kappa^2}R-\frac{1}{2} (\partial_\mu \varphi)(\partial^\mu \varphi)
-V(\varphi)
\right]\,,
\label{action1}
\end{equation}
where  $V(\varphi)$ is a potential term. For tha FLRW metric given by:
\begin{equation}
ds^2=-N^2dt^2+a^2(t)d\Omega_3^2\,,
\end{equation}
where $d\Omega_3^2$ represents a maximally symmetric three-space with a constant Ricci curvature ${}^{(3)}R_{ij}=2Kg_{ij}$, where $K$ is a constant, and $N$ is the lapse function, the corresponding Lagrangian is:
\begin{equation} \label{LagEx}
L=-\frac{1}{2N}a\dot{a}^2 +\frac{1}{2N}a^3\dot{\varphi}^2-NU(a,\varphi)\,,
\end{equation}
where $U(a,\varphi)=a^3V(\varphi)-\frac{1}{2}Ka$ (while deriving the formula above it was assumed that $\kappa^2=6$). 

The Hamiltonian associated with (\ref{LagEx}) is:
\begin{equation} \label{HamEx}
H=-\frac{1}{2}\frac{\Pi_a^2}{a} +\frac{1}{2}\frac{\Pi_\varphi^2}{a^3}+U(a,\varphi)\,,
\end{equation}
where  $\Pi_a=-\frac{a\dot{a}}{N}$ and $\Pi_\varphi=\frac{a^3\dot{\varphi}}{N}$ are the conjugate momenta. By substituting appropriate operators for the conjugate momenta in accordance with the Jordan rules $\Pi_a\rightarrow -i\frac{\partial}{\partial a}$ and $\Pi_\varphi\rightarrow -i\frac{\partial}{\partial \varphi}$ one finds the standard Wheeler-DeWitt equation for the cosmological system: 
\begin{equation}
\left[\frac{1}{a^{s+1}}\frac{\partial}{\partial a}a^s\frac{\partial}{\partial a}-\frac{1}{a^3}\frac{\partial^2}{\partial \varphi^2}+2 U(a,\varphi)\right]\Phi=0\,,
\label{WDWstandard}
\end{equation}
where the number $s$ accounts for an ambiguity in factor ordering.

The corresponding lifted Lagrangian is \cite{Finn,Kan1}:
\begin{equation}
\tilde{L}=-\frac{1}{2}a\dot{a}^2 +\frac{1}{2}a^3\dot{\varphi}^2+\frac{1}{2}\frac{\dot{\chi}^2}{2U(a,\varphi)} =\frac{1}{2}\tilde{G}_{MN}\dot{X}^M\dot{X}^N\,,
\end{equation}
where 
\begin{equation}
\tilde{G}_{MN}=\mbox{diag} (-a, a^3, [2U(a,\varphi)]^{-1})
\end{equation}
 is the metric of the extended minisuperspace and $X^M=(a,\varphi, \chi)$. We notice that the Hamiltonian constraint for the lifted system:
\begin{equation}
\frac{1}{2}\tilde{G}^{MN}\tilde{P}_M\tilde{P}_N=0\,,
\label{classical}
\end{equation}
where  $\tilde{P}_M=\tilde{G}_{MN}\dot{X}^N$, exhibits conformal invariance $\tilde{G}_{MN}\rightarrow \Omega^2 \tilde{G}_{MN}$ for an arbitrary function $\Omega(X^M)$. With the gauge choice: 
\begin{equation}
G_{MN}=2U(a,\varphi)\tilde{G}_{MN}
\label{gauge}
\end{equation}
(in the case of $n$ dimensions one should take  $G_{MN}=[2U(a,\varphi)]^{\frac{1}{n-2}}\tilde{G}_{MN}$) the extended Wheeler-DeWitt equation can be obtained by using the covariant Laplace-Beltrami operator on the extended minisuperspace which reads:
\begin{equation}
\frac{1}{\sqrt{-{G}}}\partial_M \left[ \sqrt{-{G}}{G}^{MN} \partial_N \Phi  \right]=0\,.
\label{WDWExt1}
\end{equation}
By imposing the condition $-\frac{\partial^2}{\partial \chi^2}\Phi=\Phi$ the extended Wheeler-DeWitt equation (\ref{WDWExt1})  can be reduced to the standard Wheeler-DeWitt equation given by (\ref{WDWstandard}) \cite{Kan1}.

\subsection{The extended minisuperspace Dirac-like equation}
Previous studies that aimed to take the square root of the Wheeler-DeWitt equation exhibited arbitrariness in handling the potential term \cite{Mallett,Kim,Death,Yamazaki}. The idea here is to apply the covariance principle in the extended minisuperspace, similarly as in the case of the extended Wheeler-DeWitt equation, in order to establish a consistent approach for dealing with the potential term, and thus, provide a systematic approach for determining the form of the extended minisuperspace Dirac-like equation. The Dirac-like equation in the extended minisuperspace can be formulated as \cite{Kan1}: 
\be
\label{APPdirac} 
\hat{\gamma}^\alpha D_\alpha \Psi \equiv \gamma^A \tensor{e}{_A ^\alpha} D_\alpha \Psi=0,
\ee
where the vital role is played here by the extended minisuperspace metric $G_{MN}$ in a particularly chosen gauge (see (\ref{metric_ext_conf}) or  (\ref{gauge})).  It is worth emphasizing that the solutions provided by (\ref{WDWExt1}) and (\ref{APPdirac}) are physically inequivalent solutions \cite{Kan1,Kan2}.

\end{appendices}

\end{document}